\documentstyle[aps,epsfig,floats]{revtex}
\draft
\input epsf

\begin{document}
%
\twocolumn[\hsize\textwidth\columnwidth\hsize\csname
@twocolumnfalse\endcsname

\title{Constraining Quintessence with the New CMB Data}

\author{Michael Doran, Matthew Lilley and Christof Wetterich}
 
\address{Institut f\"ur Theoretische Physik der Universit\"at Heidelberg\\
  Philosophenweg 16, D-69120 Heidelberg, Germany}
 
\maketitle
 
\pagenumbering{arabic}     
\begin{abstract}
  The CMB data recently released by BOOMERANG and MAXIMA suggest that
  the anisotropy spectrum has a third peak in the range
  $800<l_3<900$. A combination of this result with constraints from
  large-scale structure permit us to differentiate between different
  quintessence models. In particular, we find that inverse power law
  models with power $\alpha >1$ are disfavoured. Models with more than
  $5\%$ quintessence before last scattering require a spectral index
  greater than 1. 
  These constraints are compared with supernovae observations.
  We also show that the CMB alone now provides strong
  evidence for an accelerating universe.
\end{abstract}
%
 ]
 
\renewcommand{\thefootnote}{\arabic{footnote}}

Two independent observations suggest that a significant part of the
energy density is homogeneously distributed over the observable
Universe: the accelerated expansion
\cite{Perlmutter:1999np,Riess:1998cb} and the mismatch between the
amount of matter in structures and the critical energy density. An
accelerated expansion implies that the energy density of the Universe
is dominated by a component with negative pressure.  The standard
negative pressure term, Einstein's cosmological constant, is plagued
by enormous fine-tuning problems \cite{zeld,Weinberg:1989cp}, making
it seem extremely unnatural. An alternative suggestion for explaining
homogeneously-distributed dark energy is quintessence -- a scalar
field with a slowly-decaying potential
\cite{Wetterich:1988fm,Peebles:1988ek}.  Quintessence can lead to a
Universe which is accelerating today, without the severe fine-tuning
of parameters, or of initial conditions
\cite{Wetterich:1988fm,Peebles:1988ek,Copeland:1998et} (a property
referred to as tracking \cite{Steinhardt:1999nw}). The field has a
time-varying equation of state, becoming dominant only recently thus
allowing processes like nucleosynthesis and structure formation to
occur unimpeded \cite{Peebles:1988ek,Ferreira:1998hj}. For a review of
quintessence and its properties, see \cite{Wang:2000fa} or
\cite{Binetruy:2000mh}. There are several different ways of
implementing quintessence, generally involving different functional
forms for the scalar field action
\cite{Wetterich:1988fm,Peebles:1988ek,Coble:1997te,Albrecht:2000rm,Hebecker:2001zb},
or couplings to matter
\cite{Wetterich:1994bg,Amendola:2001uh,Bean:2001ys}.  These different
models have common properties, such as tracking and a negative
equation of state today, but also differ in their evolution with time.
This non-genericness of quintessence makes it difficult to devise
observational tests which could detect it and even more difficult to
rule it out. Likelihood analysis involving several different types of
observation can give good constraints on a given model, but since
there is no theoretically-preferred potential we find it more
instructive to look for generic, model-independent information.  We
seek observations sensitive to the amount of dark energy at different
epochs in the history of the Universe -- in this way the differing
time evolution of different quintessence models and a cosmological
constant can be compared.

 In a recent paper a convenient model-independent framework for
quantifying the sensitivity of the Cosmic Microwave Background (CMB)
to quintessence was proposed \cite{Doran:2000jt} (see also
\cite{Huey:1999se}).  It was demonstrated that the location of the CMB
peaks depend on three dark-energy related quantities: the amounts of
dark energy today $\Omega^0_{\phi}$ and at last scattering
$\overline{\Omega}_{\phi}^{\rm _{\, ls}}$ as well as its time-averaged
equation of state $\overline{w}_0$.  In this way, it could be possible
to extract information on the amount of quintessence present before
last scattering: if $\overline{\Omega}_{\phi}^{\rm _{\, ls}}$ turns
out to be non-zero, we would have strong evidence for non-cosmological
constant dark energy.  This procedure can also be used to
differentiate between different quintessence models.  It was
emphasized that the acoustic scale $l_A$ (which is defined below) is a
convenient single quantity for characterizing aspects of the CMB, in
the way that $\sigma_8$ (the {\em rms} mass fluctuation on scales of
$8h^{-1}$Mpc) is used for cluster abundance constraints.

Recent measurements of the CMB \cite{Netterfield:2001yq,Lee:2001yp}
show three peaks as distinct features, seeming to confirm beyond any
reasonable doubt the inflationary picture of structure formation from
predominantly adiabatic initial conditions.  In this letter we analyse
the new data and in particular the consequences of the measured peak
locations for quintessence.  We find that when combined with
constraints from large-scale structure (LSS), models where the scalar
field has an inverse-power potential are disfavoured, as are models
with more than $5\%$ quintessence before last scattering unless the
spectral index $n>1$. 
We also show that the new CMB data provides
strong evidence for an accelerating universe, independent of
supernovae (SNe Ia) data, to which we return at the end of this note.

In this work, 
we have assumed a flat universe, with
$\Omega_b h^2 = 0.022 \pm 0.003$ and $n=1$ unless otherwise stated.

The CMB peaks arise from acoustic oscillations of the primeval plasma
just before the universe becomes translucent.  The angular momentum
scale of the oscillations is set by the \emph{acoustic scale} $l_A$
which for a flat universe is given by
\begin{equation}
\label{platz}
l_A = \pi \frac{\tau_0 - \tau_{\rm ls}}{\bar c_s \tau_{\rm ls}},
\end{equation}
where $\tau_0$ and $\tau_{\rm ls}$ are the conformal time today and at
last scattering and $\bar{c}_s$ is the average sound speed before
decoupling.  The value of $l_A$ can be calculated simply, and for flat
universes is given by \cite{Doran:2000jt}

\begin{eqnarray}
\label{sep}
\nonumber l_A = \pi \bar c_s^{-1} \Bigg[
      \frac{F(\Omega^0_{\phi},\overline{w}_0)}{(1-{\overline{\Omega}^{\rm
      _{\, ls}}_{\phi}})^{1/2}} \Bigg \{ \left({a_{\rm ls} +
      \frac{\Omega^0_{\rm r}}{ 1 - \Omega^0_{\phi}}}\right)^{1/2} \\ -
      \left({\frac{\Omega^0_{\rm r}}{1 -
      \Omega^0_{\phi}}}\right)^{1/2} \Bigg \} ^{-1} - 1 \Bigg],
\end{eqnarray}  
 
with
\begin{eqnarray} \label{F_int}
\nonumber F(\Omega_{\phi}^0,\overline{w}_0) = \frac{1}{2} \int_0^1
\textrm{d}a \Bigg( a + \frac{\Omega_{\phi}^0}{1-\Omega_{\phi}^0} \, a ^{(1
- 3 \overline{w}_0)} \\ + \frac{\Omega_{\rm r}^0(1-a)}{1-\Omega_{\phi}^0}
\Bigg)^{-1/2}.
\end{eqnarray} 
Here $\Omega^{0}_{\rm r}, \Omega^{0}_{\phi}$ are today's radiation
and quintessence components, $a_{\rm ls}$ is the scale factor at last
scattering (if $a_0=1$), $\bar c_s,\overline{\Omega}^{\rm _{\,
ls}}_{\phi} $ are the average sound speed and quintessence components
before last scattering and $\overline{w}_0$ is the
$\Omega_\phi$-weighted equation of state of the Universe ($w(\tau) = p(\tau)/\rho(\tau)$)
\begin{equation}
\label{w_eff}
\overline{w}_0 = \int_0^{\tau_0} \Omega_{\phi}(\tau) w(\tau) \textrm{d}
\tau \times \left( \int_0^{\tau_0} \Omega_{\phi}(\tau) \textrm{d} \tau
\right)^{-1}.
\end{equation}

The location of the peaks is slightly shifted by driving effects and
we compensate for this by parameterising the location of the $m$-th
peak $l_m$ as in \cite{Hu:2001ti,Doran:2001yw}
\begin{equation} \label{our_phi}
l_m \equiv l_A \left(m - \varphi_m\right). 
\end{equation}
The reason for this parameterization is that the phase shifts
$\varphi_m$ of the peaks are determined predominantly by
pre-recombination physics, and are independent of the geometry of the
Universe. The values of the phase shifts are typically in the range
$0.1 \dots 0.5$ and depend on the cosmological parameters $\Omega_b
h^2, n, \overline{\Omega}^{\rm _{\, ls} }_{\phi}$ and the ratio of
radiation to matter at last scattering $r_\star = \rho_r(z_\star) /
\rho_m(z_\star).$ It is not in general possible to derive analytically
a relation between the cosmological parameters and the peak shifts,
but fitting formulae, describing their dependence on these parameters
were given in \cite{Doran:2001yw}.

\begin{figure}
 \begin{center}
 \epsfig{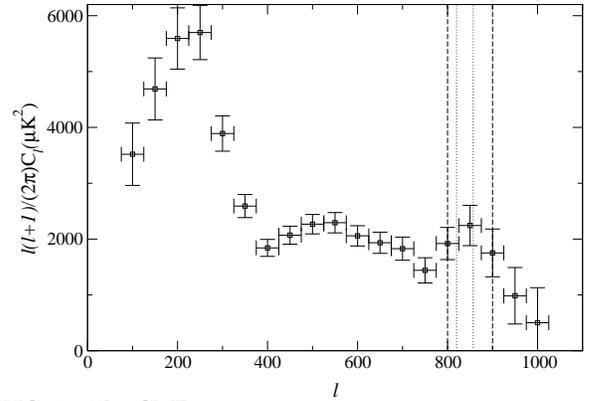} \caption{The CMB
   anisotropy power spectrum as measured by BOOMERANG [20].   
   The inner vertical lines show the
   region $820<l_3<857$ as calculated by the BOOMERANG team [24],
   and the outer lines our more conservative region
   $800<l_3<900$.\label{fig:boom}}
 \end{center}
 \end{figure}

It was shown \cite{Doran:2001yw} that $\varphi_3$ is relatively
insensitive to cosmological parameters, and that by assuming the
constant value $\varphi_3 = 0.341$ we can estimate $l_A$ to within one
percent if the location of the third peak $l_3$ is measured, via the
relation
\begin{equation}
l_A = \frac{l_3}{3 - \varphi_3}.
\end{equation}
The measurement of a third peak in the CMB spectrum by BOOMERANG
\cite{Netterfield:2001yq} now allows us to extract the acoustic scale
$l_A$ and use this as a constraint on cosmological models. The
BOOMERANG team recently performed a model-independent analysis of
their data \cite{boom}, and found the third peak to lie in the region
\begin{equation}
l_3 = 845^{+12}_{-25},
\end{equation}
from which we calculate the value
\begin{equation}\label{la_boom}
l_A = 316\pm 8.
\end{equation}
If we instead chose the more conservative assumption that
$800<l_3<900$, we would get the bound
\begin{equation}\label{la_cons}
l_A = 319 \pm 23,
\end{equation}
We will perform our analysis using both of these ranges for the
location of the third peak. The two ranges are displayed, along with
the BOOMERANG data, in Fig. \ref{fig:boom}.  Independently of
\cite{boom} we have performed cubic spline fittings to the data
presented in \cite{Netterfield:2001yq}, as well as to the combined
multiple-experiment data given in \cite{wang}.  We allowed the data to
vary according to the gaussian errors given. We find for the BOOMERANG
and combined data respectively:
\begin{eqnarray}
l_1 &=& 221 \pm 14, \quad 222 \pm 14\\
l_2 &=& 524 \pm 35, \quad 539 \pm 21\\
l_3 &=& 850 \pm 28, \quad 851 \pm 31
\end{eqnarray}   

 \begin{figure*}[ht!]
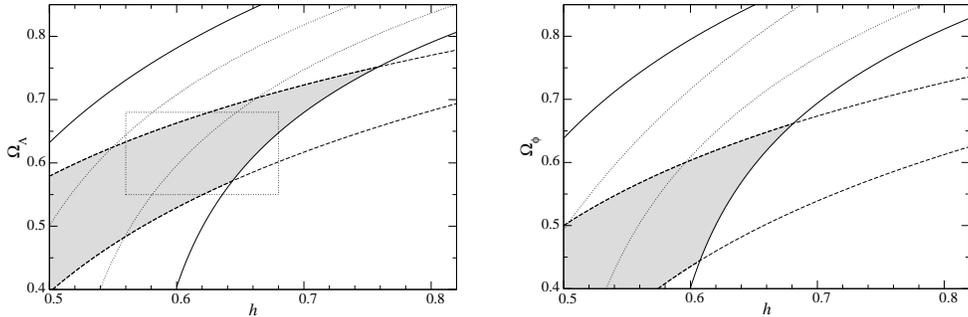

 \begin{center}
   \epsfig{file=lambda.eps ,height=6cm, angle=270} \hskip 20pt
    \epsfig{file=leaping_5.eps ,height=6cm, angle=270} 
 \caption{BOOMERANG (solid lines give conservative bound, dotted lines
    more strict bound) and LSS (dashed lines) constraints in
$\Omega_\Lambda$-$h$ plane (left) and $\Omega_\phi$-$h$ plane for LKT
quintessence with $\Omega_{\phi}^{\rm ls} = 0.05$ (right).  The dotted
box indicates the $1$-$\sigma$ maximum likelihood ranges obtained by the
BOOMERANG data analysis team with flatness and LSS
priors. \label{fig:lambda}}\end{center}
 \end{figure*}

\begin{figure*}[ht!]
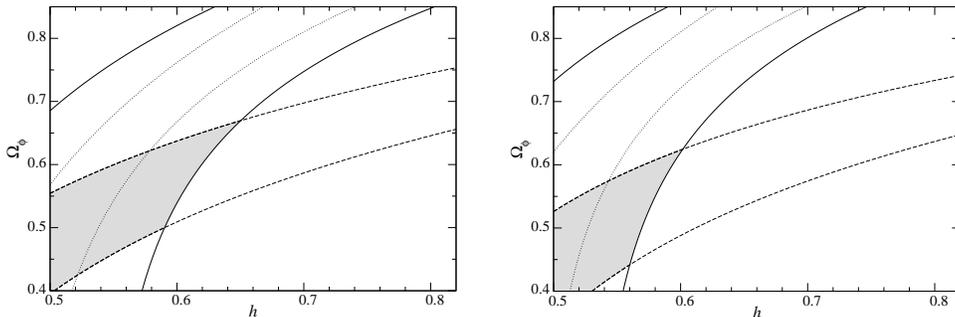

\begin{center}
  \epsfig{file=alph_1.eps,height=6cm, angle=270}\hskip 20pt
  \epsfig{file=alph_2.eps ,height=6cm, angle=270}
\caption{Constraints in the $\Omega_\phi$-$h$ plane for IPL
  quintessence, from BOOMERANG and LSS , $\alpha=1$ (left) and
$\alpha=2$ (right).
\label{fig:ipl}}
\end{center}
\end{figure*}

\begin{figure}
 \begin{center}
   \includegraphics[angle=-90,width=7.6cm]{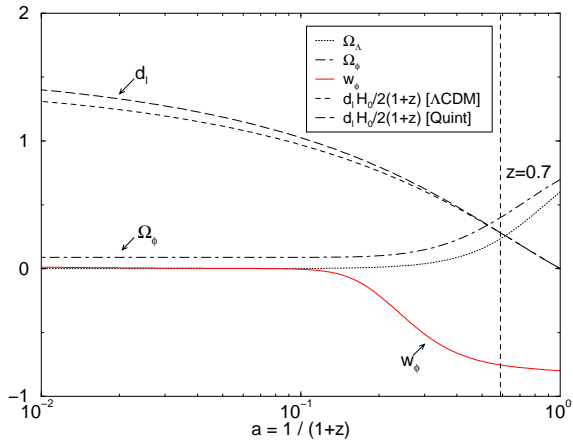} 
\caption{       
The luminosity distance $d_l(z)$ (plotted as $d_l(z) H_0 / 2(1+z)$)
and $\Omega(z)$ for a $\Lambda$CDM and a LKT universe with
$\Omega_\Lambda^0 = 0.6$ and $\Omega^0_\phi=0.7$ respectively. 
The equation
of state $w_\phi(z)$ of the LKT quintessence is also given. 
For low redshift, the
equation of state is close to $-1$, $w_0 = -0.8$.
For  $w_0[\Omega^0_\phi]^{1.4} = \Omega_\Lambda^0$, the
luminosity distance of both LKT and $\Lambda$CDM fall on top of each other  in the
redshift region relevant for current SN Ia analysis (two upper most curves). Despite
the similar late time behaviour, the LKT model has $\Omega_\phi \approx 0.1$
from very early times on, whereas in the cosmological constant
model, dark energy plays a role only recently.
}
 \label{fig:distance}
 \end{center}
 \end{figure}

We applied our CMB-derived $l_A$ constraints to two types of
quintessence model: an inverse power law (IPL) potential
\cite{Peebles:1988ek}, given by
\begin{equation}
V(\phi) = V_0 \phi^{-\alpha},
\end{equation}
and a `leaping kinetic term' (LKT) model \cite{Hebecker:2001zb}, where
the Lagrangian is given by
\begin{equation}
    {\mathcal L}(\phi) = {1\over 2}\left(\partial_\mu \phi\right)^2
    k^2\left(\phi\right) + M_{\bar{P}}^4 \exp(- \phi/M_{\bar{P}}),
\end{equation}
and kinetic term
\begin{equation}\label{tanh}
k\left(\phi\right) = k_{\rm min} + \tanh\left[\left(\phi -
\phi_1\right)/M_{\bar{P}}\right] + 1,
\end{equation} 
with $M_{\bar{P}}^{-2} = 8\pi G$.  The constants $V_0$ and $\phi_1$
determine the value of $\Omega_\phi$ today in each case. The IPL model
has equation of state today given by $w = -2/(\alpha+2)$ and in the
LKT model the constant $k_{\rm min}$ can be tuned to give specific
values of $\overline{\Omega}^{\rm _{\, ls}}_{\phi}$. 
In addition, one could multiply the argument of the $\tanh()$ in Equation
(\ref{tanh}) by a factor in order to steepen the increase in the
kinetic term. 
The equation of state today, $w_0 \equiv w(\textrm{today})$, depends 
strongly on the precise shape of $k(\phi)$. This is relevant
for supernovae observations, and we emphasize that, in general,
$w_0 \neq \overline{w}_0$. For a steep increase in $k(\phi)$, one can have
$w_0$ very close to $-1$ 
(see also figure \ref{fig:distance}).
Other models of quintessence share the effective time dependence of 
$w$ \cite{Brax:1999gp,Brax:2001ah}.

 We also applied
the constraints to a cosmological constant $(\Omega_{\phi}^0 \equiv
\Omega_{\Lambda})$ universe (i.e. IPL quintessence with $\alpha=0$)
for comparison.

In Figs \ref{fig:lambda}, \ref{fig:ipl} we show for our chosen dark
energy models the range of $\Omega_\phi$ and $h$ allowed by Equations
(\ref{la_boom}) and (\ref{la_cons}).  These ranges are similar for the
cosmological constant, LKT (also for $\overline{\Omega}^{\rm _{\,
ls}}_{\phi}=0.2$) and IPL for small $\alpha$ whereas IPL with
$\alpha=2$ would be pushed to small values of $h$.  The comparatively
low values of $h$ inferred from the BOOMERANG data can be combined
with information from LSS formation. The growth of density
fluctuations ceases when quintessence starts to dominate. In this way
LSS can serve as a probe of quintessence at intermediate
redshifts. Cluster abundance constraints for quintessence models with
constant equation of state yield \cite{Wang:1998gt}
\begin{equation}
\label{eqn::wang}
\sigma_8 \Omega_{\rm m}^{\gamma} = 0.5 - 0.1 \left[ (n-1) + (h-0.65)
\right]
\end{equation}
where $\gamma$ depends slightly on $w$, and typically $\gamma \sim
0.6$. In \cite{Wang:1998gt}, the uncertainty for Equation
(\ref{eqn::wang}) was estimated as $20 \%$ at $2$-$\sigma$, and this
is the constraint shown in the plots.  We have chosen to shade the
2-$\sigma$ LSS and conservative $l_A$ concordance region in the
$\Omega_{\phi}^0$-$h$ plane, but not to impose any bounds on these
parameters.  Recently, however, the HST has measured $h=72 \pm8$
\cite{hst}, and the 2dF survey $\Omega_m h = 0.20 \pm 0.03$
\cite{2df}.

The current CMB and LSS data are consistent with a cosmological
constant (Fig. \ref{fig:lambda}). The LKT model with $5\%$
quintessence at last scattering is marginally compatible for small
$h$.  If the amount of quintessence at last scattering is increased
beyond $5\%$, the $l_A$ bounds do not change
significantly. Compatibility with LSS data would require, however,
even higher $h$-values, at odds with the BOOMERANG data.  In contrast
to the CMB measurements, the determination of $\sigma_8$ by cluster
abundances involves systematic uncertainties that are difficult to
quantify.  Furthermore, the theoretical expectation for $\sigma_8$
depends strongly on the spectral index $n$.  Some inflationary models
indeed connect the smallness of primordial density fluctuations to
$n=1.1$--$1.15$ \cite{Wetterich:1989hf}.  Increasing $n$ increases the
amount of dark energy allowed during structure formation. For $n=1.1$,
the LKT model with $10\%$ quintessence at last scattering becomes
feasible.

The IPL model (Fig. \ref{fig:ipl}) with $\alpha = 2$ is disfavoured,
with higher values of $\alpha$ even worse, but $\alpha=1$ survives.

Of course IPL models with $\alpha < 1$ provide a better fit to the
data, however for $\alpha \rightarrow 0$ IPL approaches the
cosmological constant and the problem of naturalness becomes more and
more severe (with possible exceptions \cite{delaMacorra:2001ay}).  
Similar conclusions on the IPL model have been derived
from the old BOOMERANG data \cite{Balbi:2001kj}, but only for fixed
$h=0.65$. We see from our figures that the results can be very
sensitive to changes in $h$.

Other constraints on dark energy come from SN Ia analysis 
\cite{Perlmutter:1999jt,Huterer:2001mj,Wang:2001ht,Corasaniti:2001mf,Bean:2001xy,delaMacorra:2001xx}.
A  cosmological constant is restricted to 
$\Omega_\Lambda\in [0.5,0.9]$ at $2\sigma$ confidence level
\cite{Perlmutter:1999np,Efstathiou:1998qr}.  For quintessence, the bound on
$\Omega_\Lambda$ can easily be translated into one on $w_0$  and $\Omega^0_{\phi}$. This is due
to a degeneracy of the luminosity distance $d_l(z)$ in $w_0$ and
$\Omega^0_{\phi}$, and the fact that most of the current SNe Ia data
is in the redshift range $z \in [0.35,0.7]$.  In this range, an
approximate linear relation $d_l(z) H_0 / (1+z) = g_0(z) + x g_1(z)$ 
holds, depending only on the combination $x \equiv w_0 [\Omega^0_{\phi}]^{1.4}$.
Put another way, any Quintessence model with 
$w_0 [\Omega^0_{\phi}]^{1.4}  =-\Omega_\Lambda^{1.4}$ 
is, by current SN Ia data, indistinguishable from the corresponding $\Lambda$CDM
universe with $\Omega_\Lambda$ (see also figure \ref{fig:distance}).  
From the bounds $\Omega_\Lambda \in [0.5,0.9]$, we get
  \begin{equation}\label{sn1a} 
        -0.86\left [\Omega^0_{\phi}\right]^{-1.4} < w_0
        < -0.38 \left [\Omega^0_{\phi}\right]^{-1.4}.
  \end{equation}
For the IPL model, this can be translated into
$\Omega_\phi^0 > 0.3 (\alpha +2)^{5/7}$,  i.e. assuming 
that $\Omega_\phi^0 < 0.8$, we have $\alpha < 1.9$ (see also \cite{Corasaniti:2001mf}). 
This is comparable to our CMB and LSS constraint.
On the other hand, LKT models can be consistent with 
SNe Ia  and nevertheless differ substantially 
 from cosmological      
constant scenarios for the CMB and LSS (see figure \ref{fig:distance}). 
For these models, the CMB+LSS and the SNe Ia constraints
are not directly related and cannot easily be compared.
 \begin{figure}
 \begin{center}
   \epsfig{file=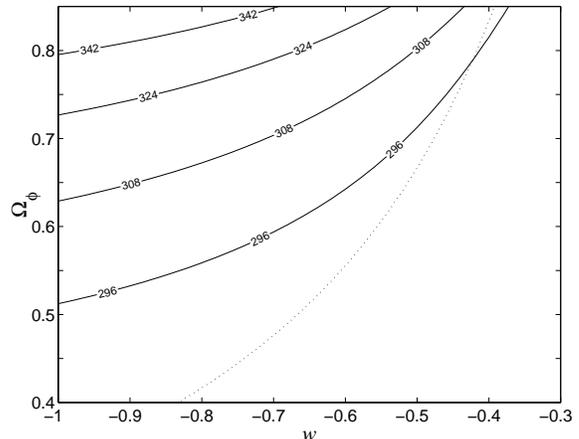 ,width=7.6cm} \caption{Lines of constant
   $l_A$ in the $\Omega_{\rm \phi}^0$-$\bar{w}_{0}$ plane, for $h=0.6$.
   All universes to the left of the dotted line are accelerating. For
   larger values of $h$, the $l_A$ lines are shifted
   north-west.\label{fig:accel}}
 \end{center}
 \end{figure}

A flat universe is accelerating today if the dark energy component and
its equation of state satisfy
\begin{equation}
\Omega_{\phi}^0 \, w_{0} < -{1\over 3}.
\end{equation}
Assuming that there is no significant dark energy component at last
scattering, we can combine our constraints on $l_A$ with Equation
(\ref{sep}).  Fig \ref{fig:accel} shows that provided $h>0.6$, the CMB
now gives strong evidence for an accelerating universe, independently
of supernovae data.

In this letter we have applied the latest CMB data to different models
of quintessence, via the easy-to-extract acoustic scale $l_A$ and
combined it with constraints from LSS formation.  We have found that
inverse power law quintessence models are severely constrained, as are
models with more than $5\%$ quintessence at last scattering and
spectral index $n=1$. In both cases the models can be compatible with
CMB or LSS when taken alone, but not together.  In order to use the
CMB to {\em detect} quintessence, via a non-zero density at last
scattering, a more accurate measurement of the location of the first
CMB peak, and hence the $\overline{\Omega}^{\rm _{\, ls}}_{\phi}$-dependent
peak shift $\varphi_1$, is required.

\section*{Acknowledgments}

We would like to thank L. Amendola, R. Crittenden, B. Mason and
J. Schwindt for helpful discussions.

\end{document}